\documentclass[a4paper,11pt]{revtex4}

\usepackage{graphicx}
\usepackage{amssymb}

\usepackage{dcolumn}

\begin{document}

\title{Net-baryon-, net-proton-, and net-charge kurtosis in heavy-ion collisions within
a relativistic transport approach}

\author{Marlene Nahrgang}
\affiliation{SUBATECH, UMR 6457, Universit\'e de Nantes, Ecole des Mines de Nantes,
IN2P3/CRNS. 4 rue Alfred Kastler, 44307 Nantes cedex 3, France}
\affiliation{Frankfurt Institute for Advanced Studies (FIAS), Ruth-Moufang-Str.~1, D-60438 Frankfurt am Main, Germany}

\author{Tim Schuster}
\affiliation{Frankfurt Institute for Advanced Studies (FIAS), Ruth-Moufang-Str.~1, D-60438 Frankfurt am Main, Germany}
\affiliation{Institut f\"ur Kernphysik, Johann Wolfgang Goethe-Universit\"at, Max von Laue-Str.~1, D-60438 Frankfurt am Main, Germany}

\author{Michael Mitrovski}
\affiliation{Institut f\"ur Theoretische Physik, Johann Wolfgang Goethe-Universit\"at, Max-von-Laue-Str.~1,
D-60438 Frankfurt am Main, Germany}
\affiliation{Frankfurt Institute for Advanced Studies (FIAS), Ruth-Moufang-Str.~1, D-60438 Frankfurt am Main, Germany}

\author{Reinhard Stock}
\affiliation{Frankfurt Institute for Advanced Studies (FIAS), Ruth-Moufang-Str.~1, D-60438 Frankfurt am Main, Germany}
\affiliation{Institut f\"ur Kernphysik, Johann Wolfgang Goethe-Universit\"at, Max von Laue-Str.~1, D-60438 Frankfurt am Main, Germany}

\author{Marcus Bleicher}
\affiliation{Institut f\"ur Theoretische Physik, Johann Wolfgang Goethe-Universit\"at, Max-von-Laue-Str.~1, D-60438 Frankfurt am Main, Germany}
\affiliation{Frankfurt Institute for Advanced Studies (FIAS), Ruth-Moufang-Str.~1, D-60438 Frankfurt am Main, Germany}

\begin{abstract}
We explore the potential of net-baryon, net-proton and net-charge kurtosis measurements to investigate the properties of hot and dense matter created in relativistic heavy-ion collisions. Contrary to calculations in a grand canonical ensemble we explicitly take into account exact electric and baryon charge conservation on an event-by-event basis. This drastically limits the width of baryon fluctuations.
A simple model to account for this is to assume a grand-canonical distribution with a sharp cut-off at the tails. 
We present baseline predictions of the energy dependence of the net-baryon, net-proton and net-charge kurtosis for central ($b\leq 2.75$~fm) Pb+Pb/Au+Au collisions from $E_{lab}=2A$~GeV to $\sqrt{s_{NN}}=200$~GeV from the UrQMD model. While the net-charge kurtosis is compatible with values around zero, the net-baryon number decreases to large negative values with decreasing beam energy. The net-proton kurtosis becomes only slightly negative for low $\sqrt{s_{NN}}$.
\end{abstract}

\maketitle

\section{Introduction}
Studies of relativistic nuclear collisions aim to elucidate the phase diagram of QCD matter at high densities and temperatures.
While non-perturbative QCD on the lattice has revealed the position of the crossover transition at vanishing baryo-chemical potential $\mu_B$ in the range of $T=145-165$~MeV depending on the observable \cite{Aoki:2006we,Borsanyi:2010bp}, the nature of the phase transition at finite $\mu_B$ is not yet ultimately discovered. Recent attempts to extend lattice QCD calculations to finite $\mu_B$ have given first indications of a critical point located between $\mu_B=200$~MeV and $\mu_B=400$~MeV \cite{Fodor:2001pe,Schmidt:2008ev}. Dramatically enhanced quark number density susceptibilities have been reported, both in two and three flavor calculations \cite{Allton:2003vx,Cheng:2009zi}. One therefore expects experimentally accessible signals of the critical point to consist of enhanced fluctuation phenomena in an event-by-event analysis of relativistic A+A collisions \cite{Stephanov:1998dy,Stephanov:1999zu,Koch:2008ia}.
Most fluctuation measures proposed are related to quadratic moments of the event-by-event observables. Among them are charged particle ratio fluctuations \cite{Jeon:1999gr,Jeon:2000wg,:2008ca}, baryon number to strangeness and charge to strangeness correlations \cite{Koch:2005vg}. Higher moments and susceptibilities, such as the skewness and the kurtosis, however, become interesting as they increase with higher powers of the correlation length of the fluctuations and are thus more sensitive to critical phenomena \cite{Stephanov:2008qz,Asakawa:2009aj}. In \cite{Stephanov:2011pb} universality class arguments were given that the kurtosis should decrease or even become negative when approaching the critical region from the crossover side.
Moments of conserved charge fluctuations are also studied in effective models with applications to heavy-ion collisions \cite{Stokic:2008jh,Skokov:2010uh,Karsch:2010ck}.
Experimental results of ratios of higher moments have been reported for Au+Au collisions at top RHIC energy \cite{Aggarwal:2010wy}. These measurements are extended to the lower energy region of $5\,{\rm GeV}\le\sqrt{s_{NN}}\le 30$~GeV, which corresponds to the interval of $\mu_B$ where a QCD critical point may be expected.

In this paper we investigate the potential of the net-baryon, net-proton and net-charge kurtosis under experimental conditions and constraints explicitly taking into account electric and baryon charge conservation. Moreover, we provide baseline predictions for the energy dependence of the kurtosis of these quantities. Here the net-charge and the net-proton kurtosis are of special experimental interest. Since the isospin susceptibility remains finite at the critical point critical fluctuations in net-baryon number are reflected in the fluctuations of the net-proton number \cite{Hatta:2003wn}. Critical phenomena leading to a non-monotonic behavior on top of this baseline could be a signal for fluctuations due to a phase transition or a critical point in the energy scan experiments at RHIC and the future Facility for Antiproton and Ion Research (FAIR) at GSI.

We shall demonstrate that the fluctuation analysis of individual collision events, as modeled by the transport calculation, reveals strong effects stemming from baryon number conservation, which depend on experimental conditions (acceptance, efficiency), and on incident energy. Moreover, we show that the net-baryon kurtosis is most affected by such conservation constraints, whereas the net-charge and net-proton kurtosis stay unaffected, thus rendering them as valuable diagnostic tools. 

The numerical simulations are done within the microscopic transport model UrQMD in version 2.3 \cite{Bleicher:1999xi,Bass:1998ca,Petersen:2008kb}. For previous event-by-event studies within the same model, the reader is referred to \cite{Bleicher:1998wu,Bleicher:2000ek,Konchakovski:2005hq}. A related study of the kurtosis has been presented in \cite{Zhou:2010us,Luo:2010by}.

The suggested observables are the quadratic and quartic susceptibilities of baryons ($B$), protons ($P$) and charged particles ($Q$):
\begin{eqnarray}
\chi_2 &=& \frac{1}{VT^3}\langle \delta N^2\rangle\\
\chi_4 &=& \frac{1}{VT^3}(\langle\delta N^4\rangle-3\langle\delta N^2\rangle^2)
\end{eqnarray}
where $\langle \delta N^4\rangle=\langle (N-\overline{N})^4\rangle$ and $\langle \delta N^2\rangle=\langle (N-\overline{N})^2\rangle$ are the fourth and second central moments of the respective distribution. The ratio of the quartic to the quadratic susceptibilities represents the kurtosis $K(\delta N)$ times the variance of the distribution, a well known statistical quantity:
\begin{equation}
\frac{\chi_4}{\chi_2}=\frac{\langle\delta N^4\rangle}{\langle\delta N^2\rangle}-3\langle\delta N^2\rangle= K(\delta N)\langle\delta N^2\rangle \equiv K^{\rm eff}\;.
\end{equation}
Analyzing the effective kurtosis instead of the kurtosis itself has the advantage that it eliminates the $1/N$ behavior as expected from the central limit theorem and removes the explicit dependence on centrality. The kurtosis of a Gauss distribution is zero, $K_{\rm Gauss}=0$. In comparison a distribution with $K>0$ (leptokurtic) has more weight in the peak and tail regions, while a distribution with $K<0$ (platykurtic) has more weight in the flanks. Thus, platykurtic distributions are generally broader but have a lower peak and more narrow tails.

\section{Transport model studies}
\begin{figure}
\centering
\includegraphics[width=0.5\textwidth]{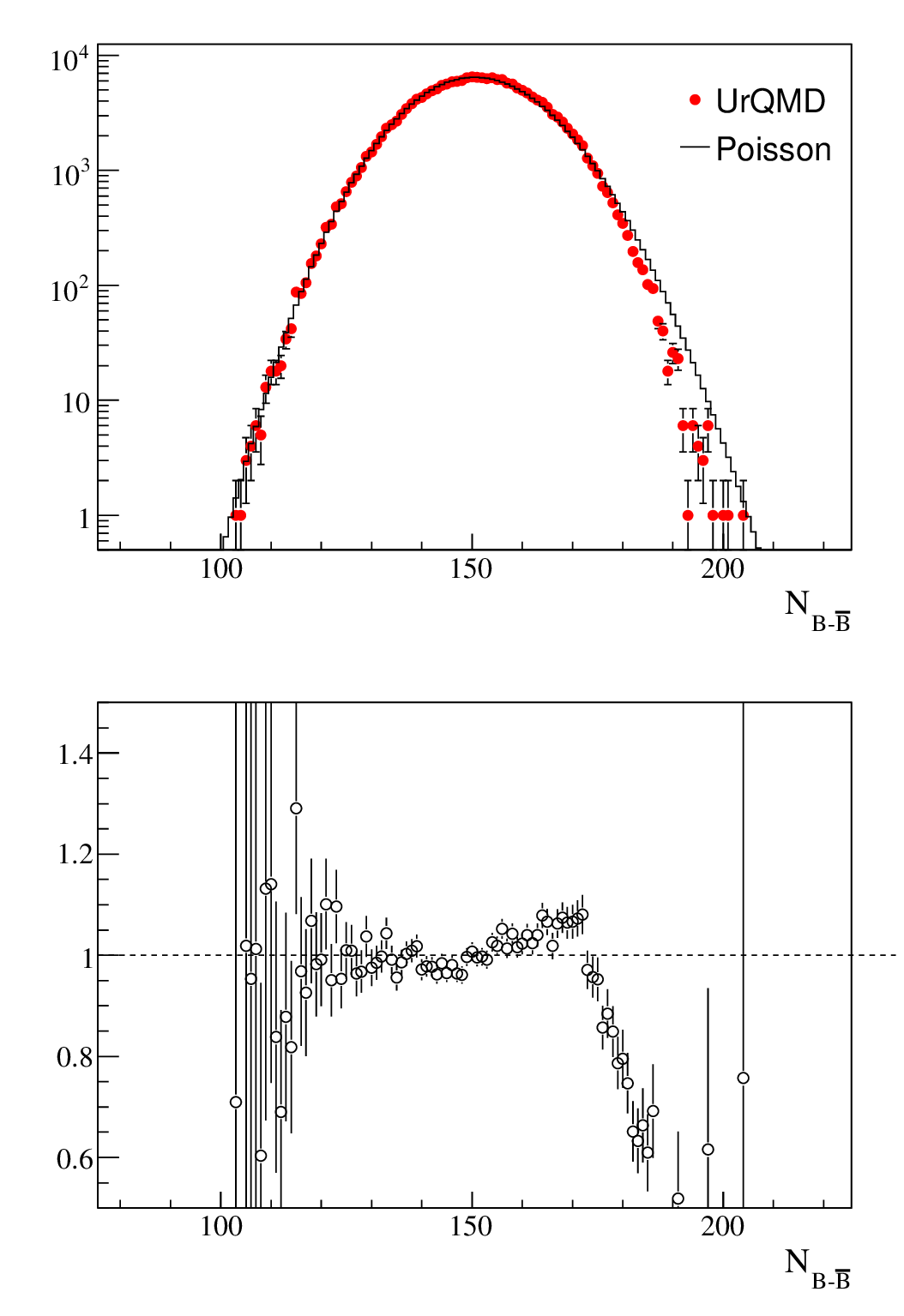}
\caption{\label{fig:figdist}Net-baryon number distribution as simulated in UrQMD for central Pb+Pb collisions at $20A$~GeV (dots) compared to a grand-canonical (Poisson) distribution with the same mean (line) and the ratio of both in the lower panel. The deviations in the shoulders and the tail are clearly visible. This leads to the decrease of the effective kurtosis from $1$ for the grand-canonical distribution to $K^{\rm eff}=-22.2$ in the data.}
\end{figure}

While the lattice QCD calculations are performed in the grand canonical ensemble, transport approaches reflect the micro canonical nature of the underlying single scattering events even in the thermodynamic limit. The choice of the thermodynamic ensemble, however, affects the fluctuations, e.g. discussed in \cite{Begun:2004gs,Cleymans:2004iu}.
In a grand canonical ensemble fluctuations of conserved charges are Poisson (or Gauss) distributed and the net charges are conserved only on average, while in real collisions net-baryon number is conserved exactly in each event. 
As a consequence the net-baryon number distribution from UrQMD simulations differs from the grand-canonical distribution at the same mean value. In Fig.~\ref{fig:figdist} we compare the simulated net-baryon number distribution for central Pb+Pb collisions at $20 A$~GeV to a grand-canonical distribution with the same mean. The acceptance window is symmetric around midrapidity ($|y|<0.5$). The lower plot of Fig.~\ref{fig:figdist} shows the ratio of the simulated distribution to the grand-canonical distribution. The simulated distribution is clearly below the grand-canonical distribution at the right tail due to the exact conservation of baryon number. Large fluctuations are suppressed. Moderate fluctuations and thus the shoulders of the simulated distribution are enhanced. It is exactly this deviation from the shape of a perfect grand-canonical distribution that leads to large negative values for the effective kurtosis obtained from the UrQMD investigations.

\begin{figure}
\centering
\includegraphics[width=0.5\textwidth]{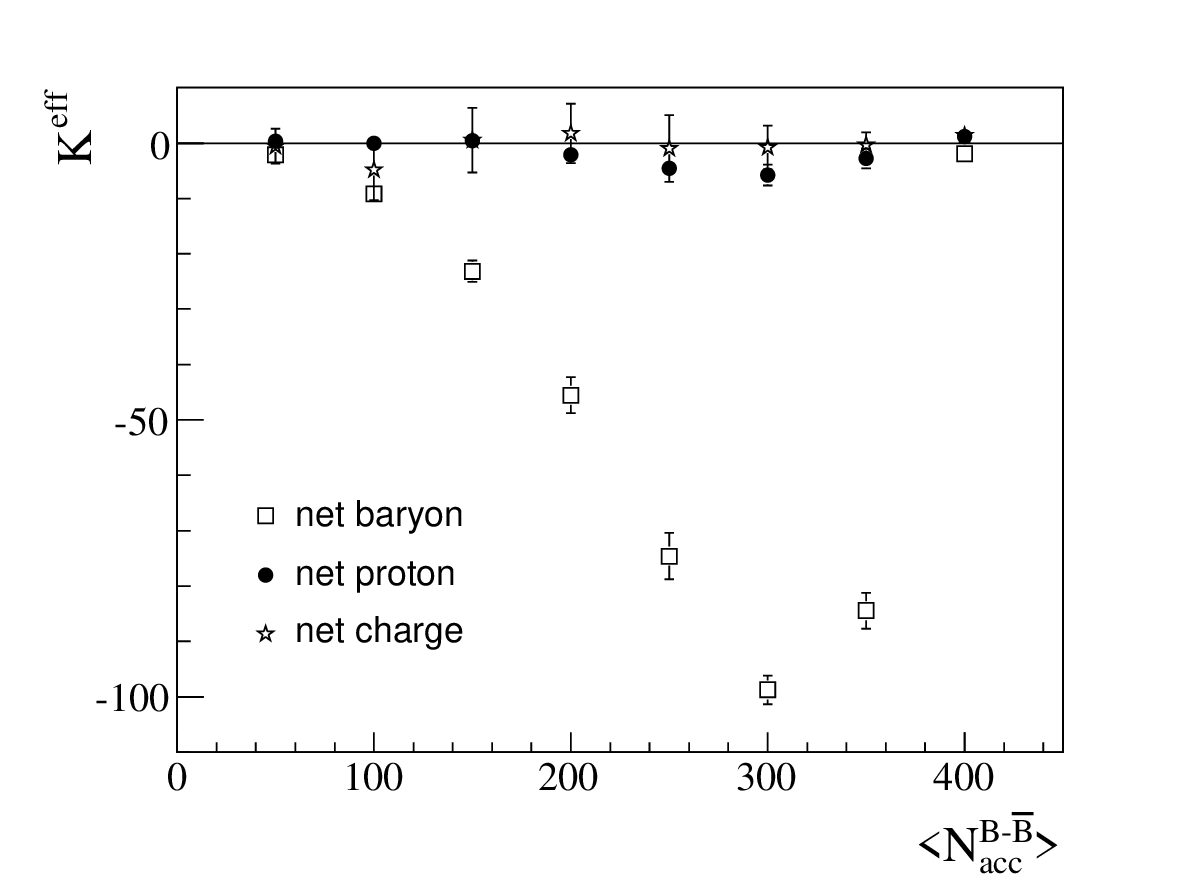}
\caption{\label{fig:fig3}Effective kurtosis of the net-charge, net-proton and net-baryon number distributions as a function of the width of the rapidity window as calculated from UrQMD for central Pb+Pb reactions at $158 A$~GeV.}
\end{figure}

In Fig.~\ref{fig:fig3}, we show the effective kurtosis of the net-charge, net-baryon and net-proton number distributions as a function of the rapidity window size in central Pb+Pb reactions at $158 A$~GeV. The latter is expressed by the mean net-baryon number in the acceptance, $\langle N^{B-\bar{B}}_{\rm acc}\rangle$. 
Broadening the rapidity window and thus going to larger mean net-baryon number leads to a decrease of the effective kurtosis until the size comes close to a $4\pi$ acceptance (in which the net-baryon number is strictly conserved, $N_{\rm tot}^{B-\bar{B}} = 416$), and the kurtosis shows a steep increase. 

\begin{figure}[t]
  \centering
  \includegraphics[width=0.5\textwidth]{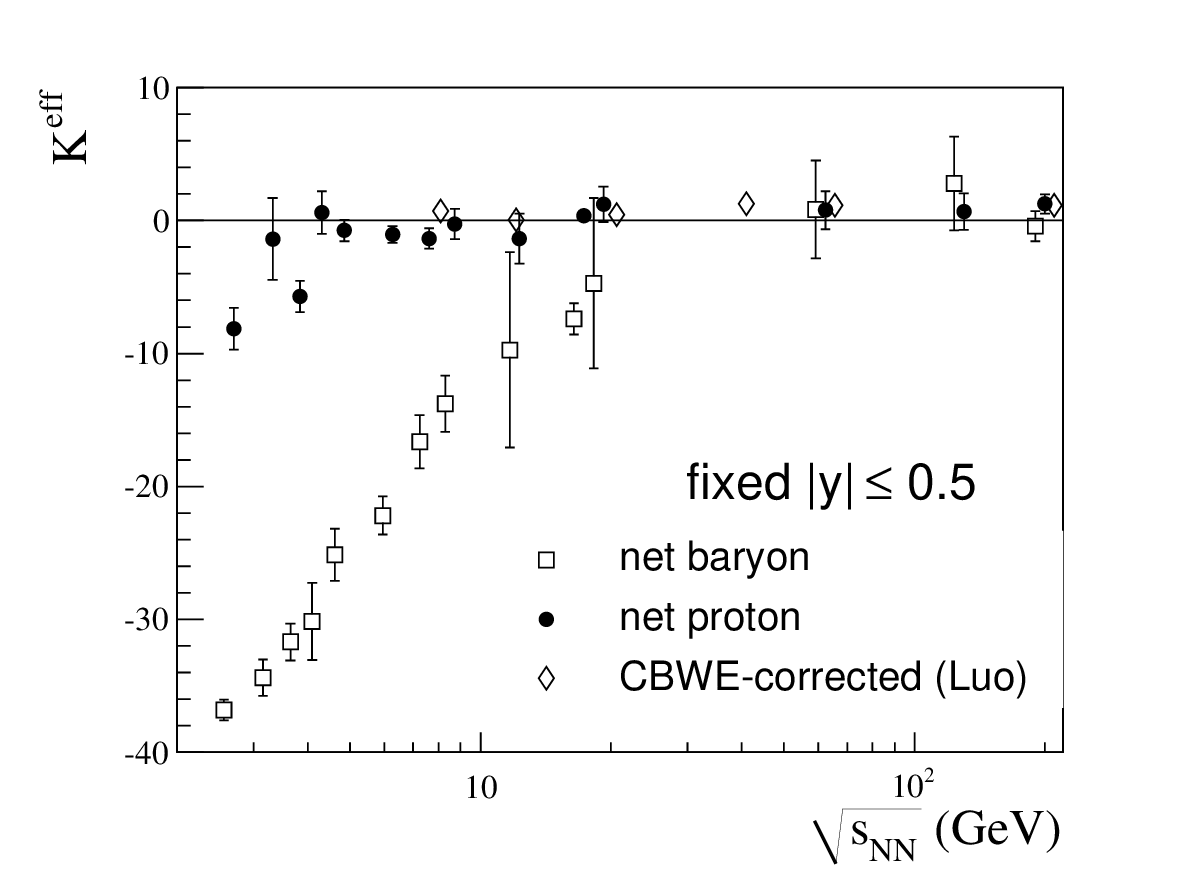}
  \caption{The energy dependence of the effective net-baryon and net-proton kurtosis at midrapidity ($|y|<0.5$) as calculated from UrQMD simulations for central ($b\leq2.75$fm) Pb+Pb/Au+Au reactions. We compare calculations without CBWE corrections to the effective net-proton kurtosis, which was corrected for CBWE \cite{Luo:2011rg}.}
  \label{fig:fig1}
 \end{figure}

The effect of baryon number conservation is also seen in the energy dependence of the effective kurtosis. When going to lower energies, the constant rapidity cut implies that a larger fraction of the total (conserved) baryon charge is observed. Fig.~\ref{fig:fig1} shows the effective kurtosis of the net-proton and net-baryon number distributions at midrapidity ($|y|<0.5$) at a range of beam energies from AGS to RHIC for central Au+Au/Pb+Pb reactions. Here one clearly observes negative values for the net-baryon number fluctuations decreasing toward lower energies. At RHIC energies the net-proton and the net-baryon number effective kurtosis is compatible with values of the order $1$ and therefore in accordance with the data from STAR \cite{Aggarwal:2010wy}. 
In both Figs., \ref{fig:fig3} and \ref{fig:fig1}, the mean numbers of the conserved charges were of the order ${\cal O}(10)$ or larger. Mean numbers of the order ${\cal O}(1)$ could be found for very small rapidity windows. These cases are not considered in the present study, the effective kurtosis for mean numbers of the order ${\cal O}(1)$ is compatible with $1$.

In \cite{Nahrgang:2011zz} we discussed the problem of superposing net-proton and net-baryon distributions with different means, which could also lead to negative values of the kurtosis and definitely affect its value. This effect could for example occur when centrality windows are chosen too broad. In the STAR analysis of the net-proton kurtosis a method to generally reduce centrality bin width effects (CBWE) is proposed  \cite{Luo:2011rg}. Here, the moments of the net-proton number distribution are calculated for each class of events with a fixed charged particle multiplicity $N_{\rm charge}$ within one centrality class. Then they are averaged by weighting with the number of events per $N_{\rm charge}$ in this centrality class. The main problem with this method is that antiprotons and protons constitute a larger fraction of all charged particles with decreasing energy. Thus, fixing $N_{\rm charge}$ puts a stronger bias on the fluctuations at lower energies. For comparison the CBWE-corrected value of the net-proton kurtosis from transport studies are shown in Fig.~\ref{fig:fig3}, too.
At energies above $\sqrt{s_{NN}}\simeq20$~GeV the uncorrected effective net-proton kurtosis is consistent with the effective kurtosis of a grand-canonical distribution and do not differ from the CBWE-corrected calculations. Differences are expected at lower energies.

\begin{figure}
\centering
\includegraphics[width=0.5\textwidth]{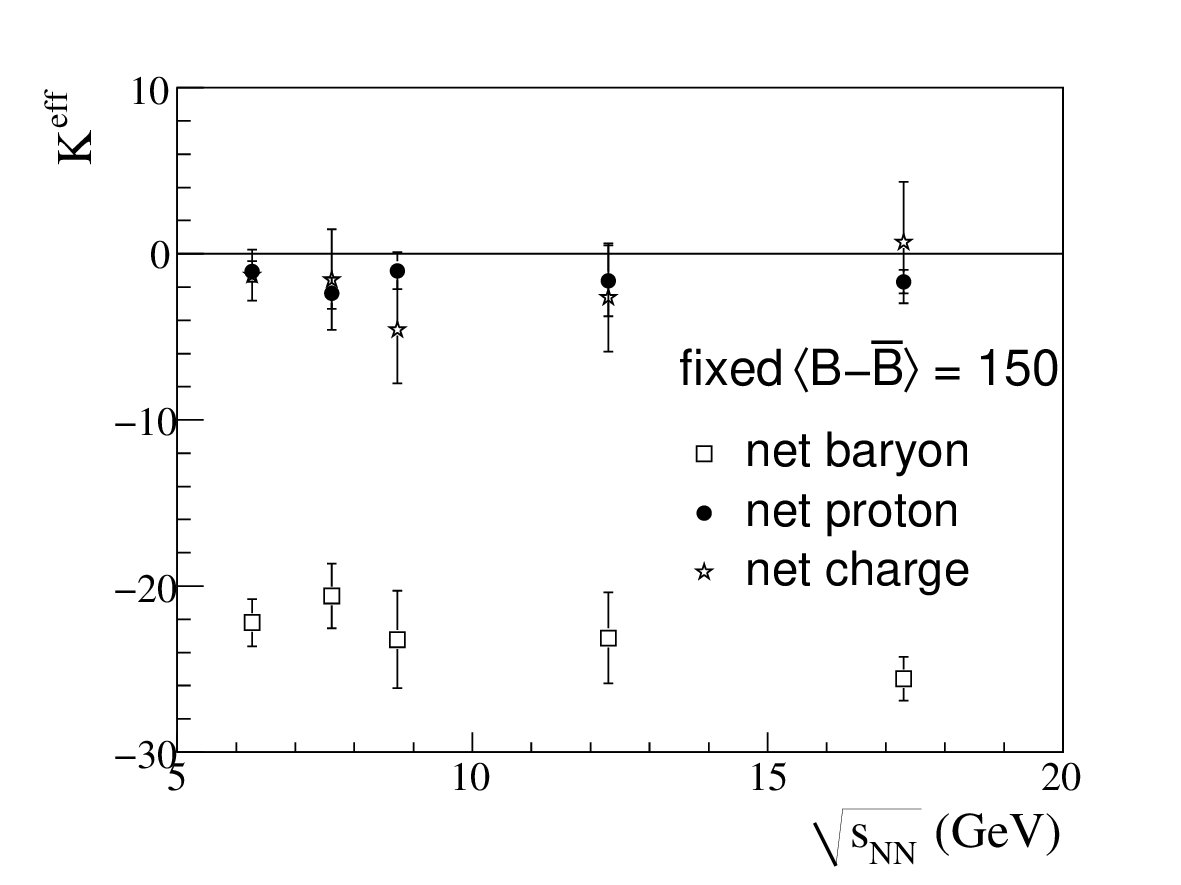}
\caption{\label{fig:a150}Effective kurtosis for net-charge, net-proton and net-baryon number distributions in acceptance windows around midrapidity covering an average net-baryon number of 150 at each energy, as calculated from UrQMD at various beam energies for central Pb+Pb reactions.}
\end{figure}

If a constant fraction of the net baryons is chosen, in contrast to a fixed rapidity window, the resulting values for the kurtosis are comparable for all investigated energies. This is illustrated in Fig.~\ref{fig:a150}, which depicts the kurtosis under selection of $150$ net baryons on average,
in Au+Au collisions. This choice corresponds to acceptances $|y|<0.6$ at $40A$~GeV and $|y|<0.85$ at $158A$~GeV, respectively.

Since protons contribute a large amount to the baryon number one should expect a similar but less pronounced effect in their distributions. However, only for lower AGS energies in Fig.~\ref{fig:fig1} the effective net-proton kurtosis becomes slightly negative resembling the constraint on net-baryon number. Also in Fig.~\ref{fig:fig3} the effective net-proton kurtosis only slightly follows the trend of the effective net-baryon kurtosis.

 The highly different sensitivity to the eventwise net-baryon number conservation constraints, as revealed by the above model calculations for net baryons and net protons, respectively, can be at least qualitatively understood by the following consideration. In collisions of Au+Au at RHIC, or Pb+Pb at the SPS, the net protons account for about 30\% of the net-baryon number. Of course, this fraction can thus fluctuate with considerably less restriction than the net-baryon number due to inelastic interactions, for example via the channels $p+n\rightarrow n+\Lambda+K^+$, or $p+n\rightarrow \Delta^++n\rightarrow \pi^++n+n$, and $p+p\rightarrow n+\Delta^{++}\rightarrow n+n+2\pi^+$.  This changes their eventwise number density in any acceptance interval, and
  lets them evade conservation constraints. These interactions do not, however, change the corresponding net baryon density which can fluctuate (in a finite acceptance) only via eventwise differences in the primordial "baryon stopping" mechanisms, or other baryonic shifts in longitudinal phase space, which also affects the fluctuations in net-proton number.  This explanation also implies that the net proton kurtosis becomes more sensitive to baryon number conservation as the average inelasticity of hadronic collisions decreases, toward low AGS energies. Fig.~\ref{fig:fig1} presents indications for this effect.

 Net charge is also a conserved quantity in a $4\pi$ acceptance but the fluctuations are less affected by this constraint. This is due to the large number of pions in a fixed rapidity window, which do not contribute to the average net charge but crucially determine the variance. Therefore net-charge fluctuations are less restricted by the constraint on the average net charge, which is determined by the protons. Even close to full acceptance charge fluctuations are still not completely suppressed. Therefore, the effective net-charge kurtosis is essentially not influenced by the conservation constraint, see in Fig.~\ref{fig:fig3}.

In RQMD \cite{Sorge:1989vt} and HIJING \cite{Gyulassy:1994ew} calculations at these energies we have also found negative values for the effective kurtosis of the same order as are observed in UrQMD calculations. 

\section{The effective kurtosis of a constrained grand-canonical distribution}
Let us illustrate how the observed deviation from a grand-canonical distribution in Fig. \ref{fig:figdist} leads to the large effective kurtosis at intermediate values of the average net-baryon number in the acceptance. We employ a simple toy model for the net-baryon number conservation and its influence on the net-baryon number distribution. The binomial distribution takes the conservation of net-baryon number in full phase-space into account. When going to large numbers the binomial distribution approaches the Poisson distribution. We want, however, to keep the range of the net-baryon number distribution finite.  We start from a Poisson distribution and modify it in the following way motivated by Fig. \ref{fig:figdist}. The narrower tail is modeled by a cut-off at the tails of the Poisson distribution and the enhancement at the shoulders is obtained from normalizing the cut distribution to unity. The model distribution is thus
\begin{equation}
P_{\mu}(N,C)={\cal N}(\mu,C) e^{-\mu}\frac{\mu^N}{N !}
\end{equation} 
on the interval $[\mu-C,\mu+C]$ and zero outside of it. $\mu$ is the expectation value of the original Poisson distribution, $C > 0$ indicates the position of the tail cut and ${\cal N}(\mu,C)$ is the normalization factor. The fluctuations of the net-baryon number in a subsystem are of the order of the square root of the average net-baryon number in this subsystem. Here, we assume that fluctuations in net-baryon number $N$ are restricted even in a comparably small rapidity window. In contrast to statistical fluctuations, an increase of the rapidity window (and therefore of the average net-baryon number in the window) does not lead to stronger fluctuations. On the contrary, when the upper limit of the total net-baryon number $N_{\rm tot}$ is approached the distribution changes to a $\delta$-function  ($K_{\delta}^{\rm eff}=0$) due to baryon number conservation. To capture this feature we thus assume a simple relation for the cut parameter
\begin{equation}
C=\alpha \sqrt{\mu}\left(1-\left(\frac{\mu}{N_{\rm tot}}\right)^2\right) .
\label{eq:cut}
\end{equation}
These values of $C$ are moderate and correspond roughly to the cut read from Fig.~\ref{fig:figdist}.
Equation (\ref{eq:cut}) describes that fluctuations are statistical for a small subsystem and vanish in the limit of full phase space coverage.

\begin{figure}
\centering
\includegraphics[width=0.5\textwidth]{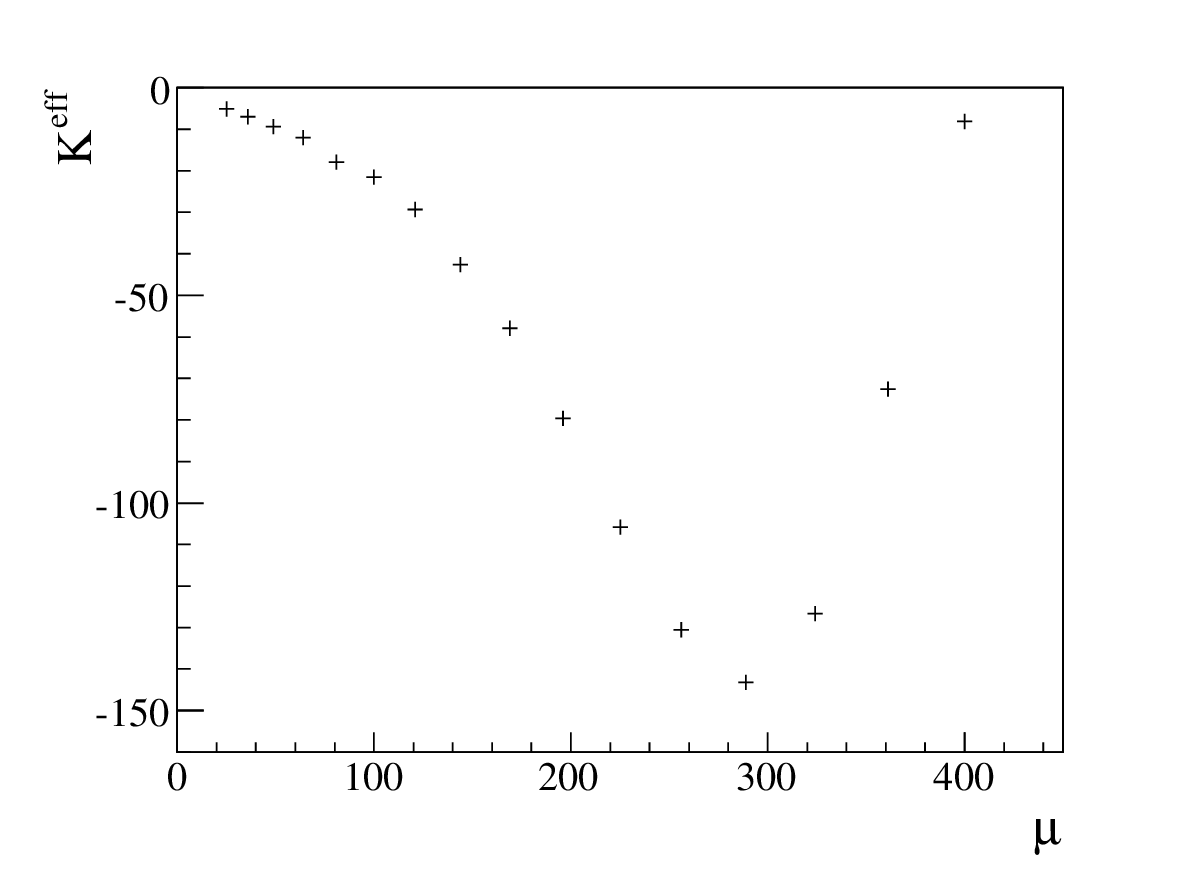}
\caption{\label{fig:plot_analytisch}Effective kurtosis as from a cut Poisson distribution. The cut parameter evolves with the expectation value $\mu$ like in equation (\ref{eq:cut}). For increasing $\mu$ the effective kurtosis first decreases until the the cut is predominantly limited by $N_{\rm tot}^{B-\bar{B}}$.}
\end{figure}

The resulting shape of the effective kurtosis is plotted in Fig. \ref{fig:plot_analytisch} with $\alpha=3$ and $N_{\rm tot}=416$. It can be seen that for this choice the effective kurtosis of the cut Poisson distribution is always negative and decreases with increasing $\mu$ until the cut affects the flanks of the distribution and weight is predominantly shifted to the peak leading to an increase of the effective kurtosis. For obtaining values of the effective kurtosis for expectation values $\mu \lesssim 10$ one would have to consider only a cut of the right tail of the distribution. As these values are not under consideration in the transport studies, we do not show them here explicitly.
The same qualitative picture can be drawn for a Gamma distribution and the kurtosis of a cut Gauss distribution is always negative.
It is remarkable how well the qualitative behavior of the effective kurtosis in Fig. \ref{fig:plot_analytisch} describes the results obtained from the UrQMD simulations in Fig. \ref{fig:fig3}. 
The quantitative values of course differ and depend on finer details such as the cut parameter and $\alpha$. 

The probability distribution of two independent Poisson-distributed random quantities is more precisely described by the Skellam distribution. A recent analytic study of the effects of including the baryon number conservation in the Skellam distribution on the cumulants of the net-proton distribution can be found in \cite{Bzdak:2012an}.

\section{Conclusions}
In summary, we have studied the energy dependence of the effective kurtosis of net-baryon, net-proton and net-charge number within a transport model of heavy-ion collisions that conserves charge and baryon number explicitly in each event. The net-baryon kurtosis was found to have large negative values below $100$~GeV. This observation, which is in contrast to the lattice results given in \cite{Cheng:2008zh} was explained by the exact baryon number conservation: the subject of the present fluctuation study. Note that the concept of an ''event`` is unknown in the grand canonical ensemble description.
A variation of the rapidity window supported this interpretation and also a simple cut in the tails of a Poisson distribution can reflect the constraint on baryon number fluctuations qualitatively. A negative kurtosis shows that moderate fluctuations are enhanced while the effect of large fluctuations is suppressed.
A realistic treatment of heavy-ion collisions at low $\sqrt{s_{NN}}$ and, thus, large baryo-chemical potential yields drastic differences as compared to the lattice QCD calculations with respect to the net-baryon kurtosis. This is especially noteworthy, because the QCD critical point is most probably located in this energy region. In our investigation, the effect of baryon number conservation alone is likely to be larger than the theoretically expected kurtosis enhancement \cite{Stephanov:2008qz} in the vicinity of a critical point of QCD. 
This is, however, different for the net-proton kurtosis and its potential to be a signal of the critical point and the first order phase transition. Down to low SPS energies the background seen in the effective net-proton kurtosis is only of the order $\pm 1$ and thus small compared to the expected signal from critical fluctuations \cite{Stephanov:2008qz,Stephanov:2011pb}. The method of correcting for centrality bin width effects as applied by the STAR collaboration has been compared to the effect of baryon number conservation in the net-proton kurtosis. At larger beam energies the two values are consistent. This method leads, however, to a bias on the flucuations at lower beam energies. These experimental methods are highly non trivial and need to be thoroughly evaluated in order to be able to extract a non-monotonic behavior of the kurtosis expected at a the critical point in energy scan programs at RHIC and FAIR.

\section*{Acknowledgements}
The authors like to thank Nu Xu and Michael Hauer for fruitful discussions. This work was supported by the Deutsche Forschungsgemeinschaft (DFG), as well as by the Hessian LOEWE initiative through HIC for FAIR. We are also grateful to the Center for Scientific Computing (CSC) at Frankfurt for providing the computing resources. M.~Nahrgang gratefully acknowledges financial support from the Stiftung Polytechnische Gesellschaft Frankfurt and T.~Schuster is grateful for support from the Helmholtz Research School on Quark Matter Studies. Moreover, this work was supported by GSI, BMBF and DESY.


\begin{thebibliography}{150}



\bibitem{Aoki:2006we}
  Y.~Aoki, G.~Endrodi, Z.~Fodor, S.~D.~Katz, K.~K.~Szabo,
  Nature {\bf 443 } (2006)  675-678.


\bibitem{Borsanyi:2010bp}
  S.~Borsanyi, Z.~Fodor, C.~Hoelbling, S.~D.~Katz, S.~Krieg, C.~Ratti and K.~K.~Szabo
                  [Wuppertal-Budapest Collaboration],
  JHEP {\bf 1009} (2010) 073.

\bibitem{Fodor:2001pe}
  Z.~Fodor and S.~D.~Katz,
  JHEP {\bf 0203} (2002) 014

\bibitem{Schmidt:2008ev}
  C.~Schmidt  [for RBC-Bielefeld Collaboration],
  J.\ Phys.\ G {\bf 35} (2008) 104093

\bibitem{Allton:2003vx}
  C.~R.~Allton, S.~Ejiri, S.~J.~Hands, O.~Kaczmarek, F.~Karsch, E.~Laermann and C.~Schmidt,
  Phys.\ Rev.\  D {\bf 68} (2003) 014507

\bibitem{Cheng:2009zi}
  M.~Cheng {\it et al.},
  Phys.\ Rev.\  D {\bf 81} (2010) 054504

\bibitem{Stephanov:1998dy}
  M.~A.~Stephanov, K.~Rajagopal and E.~V.~Shuryak,
  Phys.\ Rev.\ Lett.\  {\bf 81}, 4816 (1998)

\bibitem{Stephanov:1999zu}
  M.~A.~Stephanov, K.~Rajagopal and E.~V.~Shuryak,
  Phys.\ Rev.\  D {\bf 60}, 114028 (1999)

\bibitem{Koch:2008ia}
  V.~Koch,
  arXiv:0810.2520 [nucl-th].
  and references therein.

\bibitem{Jeon:1999gr}
  S.~Jeon and V.~Koch,
  Phys.\ Rev.\ Lett.\  {\bf 83}, 5435 (1999)

 \bibitem{Jeon:2000wg}
   S.~Jeon and V.~Koch,
  Phys.\ Rev.\ Lett.\  {\bf 85}, 2076 (2000)

\bibitem{:2008ca}
  C.~Alt {\it et al.}  [NA49 Collaboration],
  Phys.\ Rev.\  C {\bf 79} (2009) 044910



\bibitem{Koch:2005vg}
  V.~Koch, A.~Majumder and J.~Randrup,
  Phys.\ Rev.\ Lett.\  {\bf 95}, 182301 (2005)

\bibitem{Stephanov:2008qz}
  M.~A.~Stephanov,
  Phys.\ Rev.\ Lett.\  {\bf 102} (2009) 032301

\bibitem{Asakawa:2009aj}
  M.~Asakawa, S.~Ejiri and M.~Kitazawa,
  Phys.\ Rev.\ Lett.\  {\bf 103} (2009) 262301

\bibitem{Stephanov:2011pb}
  M.~A.~Stephanov,
  Phys.\ Rev.\ Lett.\  {\bf 107} (2011) 052301


\bibitem{Stokic:2008jh}
  B.~Stokic, B.~Friman and K.~Redlich,
  Phys.\ Lett.\ B {\bf 673} (2009) 192

\bibitem{Skokov:2010uh}
  V.~Skokov, B.~Friman and K.~Redlich,
  Phys.\ Rev.\ C {\bf 83} (2011) 054904

\bibitem{Karsch:2010ck}
  F.~Karsch and K.~Redlich,
  Phys.\ Lett.\ B {\bf 695} (2011) 136


\bibitem{Aggarwal:2010wy}
  M.~M.~Aggarwal {\it et al.}  [STAR Collaboration],
  Phys.\ Rev.\ Lett.\  {\bf 105} (2010) 022302

\bibitem{Luo:2011rg}
  X.~-F.~Luo, B.~Mohanty, H.~G.~Ritter, N.~Xu, B.~Mohanty, H.~G.~Ritter and N.~Xu,
  Phys.\ Atom.\ Nucl.\  {\bf 75} (2012) 676

\bibitem{Hatta:2003wn}
  Y.~Hatta and M.~A.~Stephanov,
  Phys.\ Rev.\ Lett.\  {\bf 91}, 102003 (2003)
  [Erratum-ibid.\  {\bf 91}, 129901 (2003)]



\bibitem{Bleicher:1999xi}
  M.~Bleicher {\it et al.},
  J.\ Phys.\ G {\bf 25}, 1859 (1999)

\bibitem{Bass:1998ca}
  S.~A.~Bass {\it et al.},
  Prog.\ Part.\ Nucl.\ Phys.\  {\bf 41}, 255 (1998)
  [Prog.\ Part.\ Nucl.\ Phys.\  {\bf 41}, 225 (1998)]

 \bibitem{Petersen:2008kb}
   H.~Petersen, M.~Bleicher, S.~A.~Bass and H.~Stocker,
     arXiv:0805.0567 [hep-ph].

 \bibitem{Bleicher:1998wu}
   M.~Bleicher {\it et al.},
   Phys.\ Lett.\  B {\bf 435}, 9 (1998)

 \bibitem{Bleicher:2000ek}
   M.~Bleicher, S.~Jeon and V.~Koch,
   Phys.\ Rev.\  C {\bf 62}, 061902 (2000)

 \bibitem{Konchakovski:2005hq}
   V.~P.~Konchakovski, S.~Haussler, M.~I.~Gorenstein, E.~L.~Bratkovskaya, M.~Bleicher and H.~Stoecker,
   Phys.\ Rev.\  C {\bf 73}, 034902 (2006)

\bibitem{Zhou:2010us}
  Y.~Zhou, S.~S.~Shi, K.~Xiao, K.~J.~Wu and F.~Liu,
  Phys.\ Rev.\  C {\bf 82} (2010) 014905

\bibitem{Luo:2010by}
  X.~F.~Luo, B.~Mohanty, H.~G.~Ritter and N.~Xu,
  J.\ Phys.\ G G {\bf 37} (2010) 094061


\bibitem{Begun:2004gs}
  V.~V.~Begun, M.~Gazdzicki, M.~I.~Gorenstein and O.~S.~Zozulya,
  Phys.\ Rev.\  C {\bf 70}, 034901 (2004)


\bibitem{Cleymans:2004iu}
  J.~Cleymans, K.~Redlich and L.~Turko,
  Phys.\ Rev.\ C {\bf 71} (2005) 047902


\bibitem{Nahrgang:2011zz}
  M.~Nahrgang, T.~Schuster, M.~Mitrovski, R.~Stock and M.~Bleicher,
  J.\ Phys.\ G G {\bf 38} (2011) 124150.

\bibitem{Sorge:1989vt}
  H.~Sorge, H.~Stoecker and W.~Greiner,
  Nucl.\ Phys.\  A {\bf 498} (1989) 567C.

\bibitem{Gyulassy:1994ew}
  M.~Gyulassy and X.~N.~Wang,
  Comput.\ Phys.\ Commun.\  {\bf 83} (1994) 307

\bibitem{Bzdak:2012an}
  A.~Bzdak, V.~Koch and V.~Skokov,
  arXiv:1203.4529 [hep-ph].


\bibitem{Cheng:2008zh}
  M.~Cheng {\it et al.},
  Phys.\ Rev.\  D {\bf 79} (2009) 074505

\end{thebibliography}
\end{document}